\documentclass{article}
\usepackage{arxiv}
\usepackage[utf8]{inputenc} 
\usepackage[T1]{fontenc}    
\usepackage{hyperref}       
\usepackage{url}           
\usepackage{booktabs}       
\usepackage{amsfonts}      
\usepackage{nicefrac}      
\usepackage{microtype}      
\usepackage{lipsum}	
\usepackage{graphicx}
\usepackage{tikz}
\usetikzlibrary{arrows,calc,positioning}
\usepackage{pgfplots}
\usepackage{graphicx}
\DeclareUnicodeCharacter{FB01}{fi}
\usepackage{caption}
\usepackage{csquotes}
\usepackage{colortbl}
\usepackage{multirow}
\usepackage{tabularx}
\usepackage{multicol}
\usepackage{rotating} 
\usetikzlibrary{arrows,shadows} 
\usepackage[underline=true,rounded corners=false]{pgf-umlsd}
\usepackage{enumitem}
\newlist{steps}{enumerate}{1}
\setlist[steps, 1]{label = Step \arabic*:}
\tikzstyle{intt}=[draw,text centered,minimum size=6em,text width=5.25cm,text height=0.34cm]
\tikzstyle{intl}=[draw,text centered,minimum size=2em,text width=2.75cm,text height=0.34cm]
\tikzstyle{int}=[draw,minimum size=2.5em,text centered,text width=3.5cm]
\tikzstyle{intg}=[draw,minimum size=.5em,text centered,text width=1.2cm]
\tikzstyle{intg1}=[draw,minimum size=1em,text centered,text width=1.8cm]
\tikzstyle{intg2}=[draw,minimum size=1em,text centered,text width=2.1cm]
\tikzset{%
   neuron missing/.style={
    draw=none, 
    scale=4,
    text height=0.333cm,
    execute at begin node=\color{black}$\vdots$
  },
}
\title{A Privacy-Preserving Outsourced Data Model in Cloud Environment}
\author{ \href{https://orcid.org/0000-0001-9305-1269}{\includegraphics[scale=0.06]{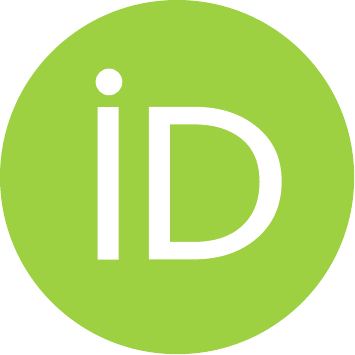}\hspace{1mm}Rishabh~Gupta}\thanks{This work is financially supported by the University Grant Commission, New Delhi, India, under the scheme of National Eligibility Test-Junior Research Fellowship (NET-JRF) with reference id-3515/(NET-NOV 2017)} \\
	Department of Computer Applications\\
	National Institute of Technology\\
	Kurukshetra, India \\
	\texttt{rishabh\_6180047@nitkkr.ac.in} \\
	\And
	\href{https://orcid.org/0000-0002-8053-5050}{\includegraphics[scale=0.06]{orcid.pdf}\hspace{1mm}Ashutosh Kumar~Singh} \\
	Department of Computer Applications\\
	National Institute of Technology\\
	Kurukshetra, India \\
	\texttt{ashutosh@nitkkr.ac.in} \\
}
\renewcommand{\shorttitle}{A Privacy-Preserving Outsourced Data Model}
\hypersetup{
pdftitle={A template for the arxiv style},
pdfsubject={q-bio.NC, q-bio.QM},
pdfauthor={Rishabh~Gupta, Ashutosh Kumar~Singh},
pdfkeywords={First keyword, Second keyword, More},
}
\begin{document}
\maketitle

\begin{abstract}
Nowadays, more and more machine learning applications, such as medical diagnosis, online fraud detection, email spam filtering, etc., services are provided by cloud computing. The cloud service provider collects the data from the various owners to train or classify the machine learning system in the cloud environment. However, multiple data owners may not entirely rely on the cloud platform that a third party engages. Therefore, data security and privacy problems are among the critical hindrances to using machine learning tools, particularly with multiple data owners. In addition, unauthorized entities can detect the statistical input data and infer the machine learning model's parameters. Therefore, a privacy-preserving model is proposed, which protects the privacy of the data without compromising machine learning efficiency. In order to protect the data of data owners, the $\epsilon$-differential privacy is used, and fog nodes are used to address the problem of the lower bandwidth and latency in this proposed scheme. The noise is produced by the $\epsilon$-differential mechanism, which is then added to the data. Moreover, the noise is injected at the data owner’s site to protect the owners' data. Fog nodes collect the noise-added data from the data owners, then shift it to the cloud platform for storage, computation, and performing the classification tasks purposes.
\end{abstract}
\section{Introduction}
Nowadays, Machine learning algorithm has become an essential tool in many applications, including product recommendations, fraud detection, speech recognition, etc. The cloud service providers such as Amazon Web Service (AWS), Google, Microsoft Azure, etc., offer ‘ML-as-a-Service’ (MLaaS) service as a black-box application programming interface (API) that accomplishes machine learning classification tasks \cite{gupta2022differentialAIHC, saxena2021survey, saxena2022sustainable}. Data owners can achieve classified results by uploading data sets and performing the tasks of classification \cite{gupta2020mlpam, gupta2022quantum} by employing the machine learning classification model \cite{singh2022metaheuristic, goel2022facial, swain2022efficient}.
To train the classification model, it requires an ample amount of data from various sources, which can contain sensitive information, including intellectual properties, personal images, and health-related records \cite{tripathihedcm, tripathi2020review, singh2020online}. These data are acquired from multiple owners \cite{yadav2019advancements, gupta2019dynamic, tiwari2021credit}. Initially, data owners shift their data to the cloud service provider at the cloud server for storage, analysis, and classification tasks \cite{chhabra2016dynamic}. This data is passed through the fog nodes. The cloud service provider accesses all data from the fog nodes and stores it on its server \cite{varshney2021machine, choudhary2021review}. When the data from the owners' site is outsourced to the cloud server, owners lose ownership of their data at that time and are unaware of the outsourced data accessing it \cite{kumar2020ensemble, chhabra2020secure, saxena2021workload, singh2021quantum, saxena2021op, kumar2020biphase}. Even outsourced data may be accessed by unauthorized entities, or any adversary \cite{singh2019secure, saxena2022intelligent, saxena2022ofp}. The data owners may not believe on the fog nodes and cloud service provider, since third parties manage it \cite{gupta2022holistic, saxena2021osc, gupta2021data, saxena2021secure, saxena2022high, saxena2022fault}. 
Although the owner’s data play an essential role in classification tasks in the machine learning model, disclosing this data will lead to some significant privacy concerns \cite{gupta2020framework, gupta2020integrated, gupta2020seli}. Therefore, to protect the data, the owners initially encrypt their data for privacy reasons before transferring it to other entities \cite{kumar2021discussion,singh2022privacy}. Nevertheless, the data is encrypted using some of the most well-known encryption techniques \cite{gupta2018probabilistic, gupta2019layer, deepika2020review, chauhan2020survey, singh2021cryptography, saxena2020security, chhabra2020security, gupta2020guim, acharya2022maci}. But, using this encrypted data with a machine-learning model is quite challenging \cite{gupta2022privacyICPC2T}. 
Another problem is that an increase in the number of internet devices may lead to a rise in demand for low-latency real-time services that will be difficult for traditional cloud computing \cite{mittal2021study, kumar2021self, kumar2020decomposition, chhabra2018oph, chhabra2019dynamic, saxena2021proactive, gupta2022privacy, chhabra2022comprehensive, saxena2021energy}. To tackle the aforementioned problems, we have proposed a model named \textbf{P}rivacy-\textbf{P}reserving \textbf{O}utsourced \textbf{D}ata (PPOD) model in cloud environment.  Our proposed scheme is based on fog computing and differential privacy. Initially, each data owner transfers his data to each fog node under this scheme. Since they don't interact with one another, different privacy protection is taken into consideration at the owner's site. According to the various applications of the classification, each data owner injects different statistical noises into outsourced data. All the fog nodes collect the noisy data from each data owner and transfer all the noise-added data to the cloud service provider whenever the data owner sends the request for the classification tasks. The classification service is provided by the cloud service provider \cite{jalwa2021comprehensive, sharma2021lightweight, gupta2022differentialJWE}.  It gets noisy data from the data owner through the fog nodes. The classification task is carried out over the data collected from the data owners. The cloud service provider sends the result of the classification to the data owner.  Therefore, our proposed scheme performs the machine learning classification task in a manner that preserves privacy. 
\section{Related Work}
\subsection{Privacy-Preserving Based on Cryptography}
In order to perform classification tasks on the ciphertext under multiple keys, Ma et al. \cite{ma2018pdlm} devised a privacy-preserving deep learning model, referred to as PDLM, that enables the service provider (SP) to migrate majority computing to the cloud to train a deep learning model based on stochastic gradient descent (SGD). It also accomplishes the feed-forward and back-propagation procedure based on an effective privacy-preserving calculation toolkit in the cloud without releasing any confidentiality. By doing this, SP's storage and computational cost are reduced while ensuring the security of the training data. The experiments were conducted, and the result indicated that the PDLM model could effectively and efficiently train model while protecting data privacy. This model's classification accuracy result is very low and has a high computation cost.
To tackle the issue of training deep neural network algorithms over encrypted data, Hesamifard et al. \cite{hesamifard2018privacy} presented a framework, namely CryptoDL, that preserves data privacy using homomorphic encryption. They provided the theoretical basis and demonstrated that it is feasible to identify the lowest degree polynomial approximation of a function within a given error bound. They developed entire neural networks over homomorphically encrypted data, with polynomial approximations used in place of the activation functions (Sigmoid and ReLU). The empirical results demonstrated that the proposed CryptoDL provided accurate privacy-preserving training and classiﬁcation. While this secures personal information well, they did not take into account the need that several keys to protect confidential data from different data owners.
A scheme for a classifier owner to delegate a remote server was introduced in \cite{li2018outsourced}, which offers the privacy-preserving classification service for users. The authors developed effective encryption methods for Naive Bayes and hyperplane decision-based classification, respectively. On the LAN server, the experiments were carried out. This system's disadvantage is that users' interactions are frequently involved when a classification query is launched.
Aono et al. \cite{aono2017privacy} devised a system called gradients encrypted asynchronous SGD (ASGD), which utilizes additively homomorphic encryption to secure the gradients across an honest-but-curious cloud server. The proposed method outperformed a similar deep learning system that had been trained on the shared dataset of all participants in accuracy terms. On the cloud server, all gradients are encrypted and kept safe.  The computation across gradients is made possible by the additive homomorphic characteristic. The proposed scheme's drawback is that the proposed scheme's trade-off is the additional communication overhead between the cloud server and DL participants.
A multiparty Back-Propagation neural (BPN) network learning scheme is presented in \cite{yuan2013privacy}, which safeguards each participant's private data set and intermediate outcomes produced throughout the network learning process. With a data set that has been arbitrarily partitioned, the proposed method enables collaborative learning between two or more parties. The BGN homomorphic encryption technique was adopted and modified to provide flexible operations over ciphertexts.  The proposed scheme has been demonstrated to be secure, effective, and accurate through numerical analysis and experiments on commodity clouds. However, the BGN cryptosystem only allows for a single multiplication operation and an infinite number of addition operations. Therefore, a fully homomorphic encryption strategy was utilized to safeguard the confidentiality of the encrypted data and allow an arbitrary amount of multiplication and addition operations.
To maintain the multiplicative depth of the neural network, Chabanne et al. \cite{chabanne2017privacy} designed and evaluated the privacy-preserving classification for neural networks with a depth greater than 2. The fundamental concept was to combine simplifications of the NN with Fully Homomorphic Encryptions (FHE) approaches to achieve both processing efficiency and data secrecy. They substituted low-degree polynomial approximations for the ReLU functions in the CNN utilized in the classification phase. The high degree is present in the ReLU function. Consequently, a batch normalization layer was added to the ReLU's polynomial approximation. Results of the experiments indicated that the proposed framework obtained private classification accuracy on the private data is better than Cryptonets and near to non-private data accuracy. Moreover, devices with limited resources cannot handle homomorphic encryption's high computational requirements. A summary of the literature review of privacy-preserving based on Cryptography is shown in Table 1.
	\begin{table}[htbp]
		\caption{Pandect Summary of Privacy-Preserving driven on Cryptography}
		\label{TableExp1}
		\begin{center}
		\small\addtolength{\tabcolsep}{-4pt}
			\begin{tabular}{|p{2.5cm}|p{2cm}|p{2cm}|p{2cm}|p{3.5cm}|p{3cm}|}\hline 
				\textbf{\textit{Methods/Schemes}} & 
				\textbf{\textit{Strategy}}& \textbf{\textit{Datasets}}& \textbf{\textit{Classifiers}}&
					\textbf{\textit{Pros}} &  
						\textbf{\textit{Cons}} \\  \hline  \hline
A secure deep learning model for training the encrypted data \cite{ma2018pdlm} & DD-PKE & MNIST, CIFAR-10 & SGD, Sigmoid Function & Trained the deep learning model efficiently and effectively & The classification accuracy result of this model is significantly less and has a high computation cost \\ \hline
A privacy-preserving model for training the neural network \cite{hesamifard2018privacy} & Homomorphic encryption and neural networks & MNIST, CIFAR-10 & Sigmoid and ReLU activation functions & Provided accurate privacy-preserving training and classification & They did not consider the requirement that private data be protected by several keys from different data owners \\ \hline
A privacy-preserving outsourcing scheme for the service of classification \cite{li2018outsourced} & Additive homomorphic encryption, Private Information Retrieve (PIR) & Balance Scale, Breast Cancer Original, SPECT Heart, Gene Sequences, Bank Marketing, Image Segmentation, Breast Cancer Diagnostic and Steel Plates Faults & Naive Bayes and the hyperplane decision & Preserved the privacy of the classification model & Communications among users continually involved when establishing a classification query \\ \hline
A secure parameters protection scheme for deep learning \cite{aono2017privacy} & Additively homomorphic encryption & MNIST, SVHN & Asynchronous Stochastic Gradient Descent & Achieved high accuracy and retained the data privacy & The trade-off increases the communication overhead \\ \hline
A secure data protection scheme for sharing cipher-text \cite{yuan2013privacy} & BGN doubly homomorphic  & Iris, kr-vs-kp, and diabetes & Back-Propagation Neural Network Learning, Sigmoid & The computation and transmission costs for each group are the least cost and are independent of the number of participants & BGN cryptosystem sustains only one multiplication but performs addition operations on multiple terms \\ \hline
A secure deep neural network for classification \cite{chabanne2017privacy} & Fully Homomorphic Encryption (FHE) & MNIST & ReLU, Sigmoid function & It provides better accuracy & The homomorphic encryption is too computation \\ \hline
			\end{tabular}
		\end{center}
	\end{table}
\subsection{Privacy-Preserving Based on Differential Privacy}
To preserve the privacy of sensitive information of the data owners for classification, Li. et al. \cite{li2018differentially} devised a privacy-preserving Naive Bayes learning scheme using multiple data sources. The proposed scheme enables a trainer to train a Naive Bayes classifier over the dataset that is jointly provided by various data owners without the assistance of a reliable curator. They designed the aggregation approach with the intention of hiding some statistical data (e.g., the number of total samples).  The experiments which were conducted on a LAN server demonstrated the viability of the proposed approach for use in various applications.
To accomplish privacy-preserving machine learning over cloud data from various data providers, Li et al. \cite{li2018privacya} proposed a scheme that protects the data sets of different providers and the cloud, respectively. The data sets of different providers have been encrypted by adopting the separate public keys by the public-key encryption with a double decryption algorithm (DD-PKE) cryptosystem. It allows the encrypted data to be transformed into a randomized data set without information leakage, while also protecting the cloud's data sets using $\epsilon$-differential privacy. The experiments demonstrated that the proposed scheme enhanced data analysis precision and computational effectiveness.
	\begin{table}[htbp]
		\caption{Pandect Summary of Privacy-Preserving driven on Differential privacy}
		\label{TableExp2}
		\begin{center}
		\small\addtolength{\tabcolsep}{-4pt}
			\begin{tabular}{|p{2.5cm}|p{2cm}|p{2cm}|p{2cm}|p{3.5cm}|p{3cm}|}\hline 
				\textbf{\textit{Methods/Schemes}} & 
				\textbf{\textit{Strategy}}& \textbf{\textit{Datasets}}& \textbf{\textit{Classifiers}}&
					\textbf{\textit{Pros}} &  
						\textbf{\textit{Cons}} \\  \hline  \hline
A secure classifier scheme for learning the process \cite{li2018differentially} & Differential privacy & Balance Scale, Breast Cancer Original, SPECH Heart, Gene Sequences & Naive Bayes & Designed Algorithm hides the statistics information and do not involve heavy cryptographic tools & In this approach, collisions are allowed, or adversaries have the ability to alter the actual data \\ \hline
A secure data sharing scheme \cite{li2018privacya} & Differential Privacy and Homomorphic encryption & Abalone, Wine, Cpu, Glass, and Krkopt & K-NN, SVM, Random Forest, and Naïve Bayes & Improved the efficiency of the computation & SVM classification based on Laplace has less accuracy \\ \hline
A secure weight-update-based framework for input privacy \cite{phuong2019privacy} & Server-aided Network Topology, Fully-connected Network Topology & MNIST, CIFAR-10, CIFAR-100, Pima (diabetes), Breast Cancer, Banknote Authentication, Adult Income, Skin/NonSkin & The stochastic gradient descent (SGD) algorithm, ReLU, Sigmoid & Low computation and communication overhead and high accuracy & Unable to parallel training, the deep learning model is sequentially trained over the datasets \\ \hline
A secure and efficient data release scheme \cite{ye2020secure} & Differential Privacy & Original & Not available & Maintaind the data integrity & Less data utility\\ \hline
Secure outsourcing scheme for data publication \cite{li2018efficient} & Differential Privacy and Additively homomorphic encryption & Letter Recognition, EEG Eye State, CPU, and Glass & KNN and Naïve Bayes & More efficiency & An efficient method does not exist in the proposed scheme \\ \hline
 A differential privacy model for sensitive data \cite{singh2022privacyMTA} & Differential Privacy  &  Heart Disease, Arrhythmia, Hepatitis, Indian-liver-patient, Framingham & SVM, Random Forest,KNN, Naïve Bayes, ANN & Achieved accuracy up to 93.75\% & It does not protect the classification model \\ \hline
A data and classification protection model \cite{gupta2022differentialNGC} & Differential Privacy & Iris, Heart disease, nursery, and balance scale & Naive Bayes & Acquired accuracy up to 94\% & Performance degradation \\ \hline
A machine learning based model for medical data \cite{gupta2022differential} & Differential Privacy & Parkinson, SpectF, Planning Relax, and Wisconsin Breast Cancer & Deep neural network & Achieved accuracy up to 87.03\% & It can be extended by devising more efficient privacy-preserving mechanism \\ \hline

			\end{tabular}
		\end{center}
	\end{table}
The Server-aided Network Topology (SNT) system and the Fully-connected Network Topology (FNT) system, depending on the connection with SNT or FNT among the trainers, were presented by Phong and Phuong \cite{phuong2019privacy} that protects the specific data of all trainers. The stochastic gradient descent (SGD) algorithm or its derivatives can be used by multiple machine-learning trainers in the proposed systems to analyze the combined datasets of the trainers without exposing the local datasets of the individual trainers. Instead of using gradient parameters, these systems used weight parameters. They preserved privacy while achieving the same learning accuracy as SGD. The experiments were carried out, and the outcomes demonstrated that the proposed systems are practically effective in terms of computation, communication, and accuracy. However, the proposed scheme is not able to train the deep learning model parallel sequentially trained on private datasets.
To address the issue of valuable data and responding thousands of queries under differential privacy (DP) is inappropriate and unprotected to attack, Ye et al. \cite{ye2020secure} proposed a secure and efficient outsourcing DP data release scheme with order-preserving encryption (OPE) rather than the general solution utilizing homomorphic encryption in the cyber-physical system. The proposed method allowed for the outsourcing of datasets by data providers to a cloud service provider with reduced connectivity costs. After uploading their encrypted data, data providers were not required to be online. Less accuracy is provided by the proposed scheme.
To address the problems of effectiveness related to adding various types of noise to a dataset for differentially private publication, An efficient and secure outsourced differential privacy scheme was introduced in \cite{li2018efficient}. Additive homomorphic encryption was employed to encrypt the data,while noise was generated by using differential privacy. The proposed scheme allowed data providers to contract with cloud service providers with low communication costs to handle their dataset sanitization process.  Furthermore, after uploading their datasets and noise parameters, data providers can go offline, which is another essential requirement for a practical system. Moreover, experiments were evaluated, confirming the proposed scheme's effectiveness. However, an efficient method only exists in the proposed system that permits data to be released once while maintaining the data's usefulness for numerous evaluation algorithms and applications is present in the proposed system. A summary of the literature review of privacy-preserving based on differential privacy is depicted  in Table 2.
\section{System Model}
The system model consists of three entities, such as Data Owners ($DO$), Fog Nodes ($FN$), and Cloud Service Provider ($CSP$). The system model is split into three-tier, as shown in Fig. 1. All the tiers are described below:
	\begin{figure*}[!htbp]
		\centering{\includegraphics[scale=.4]{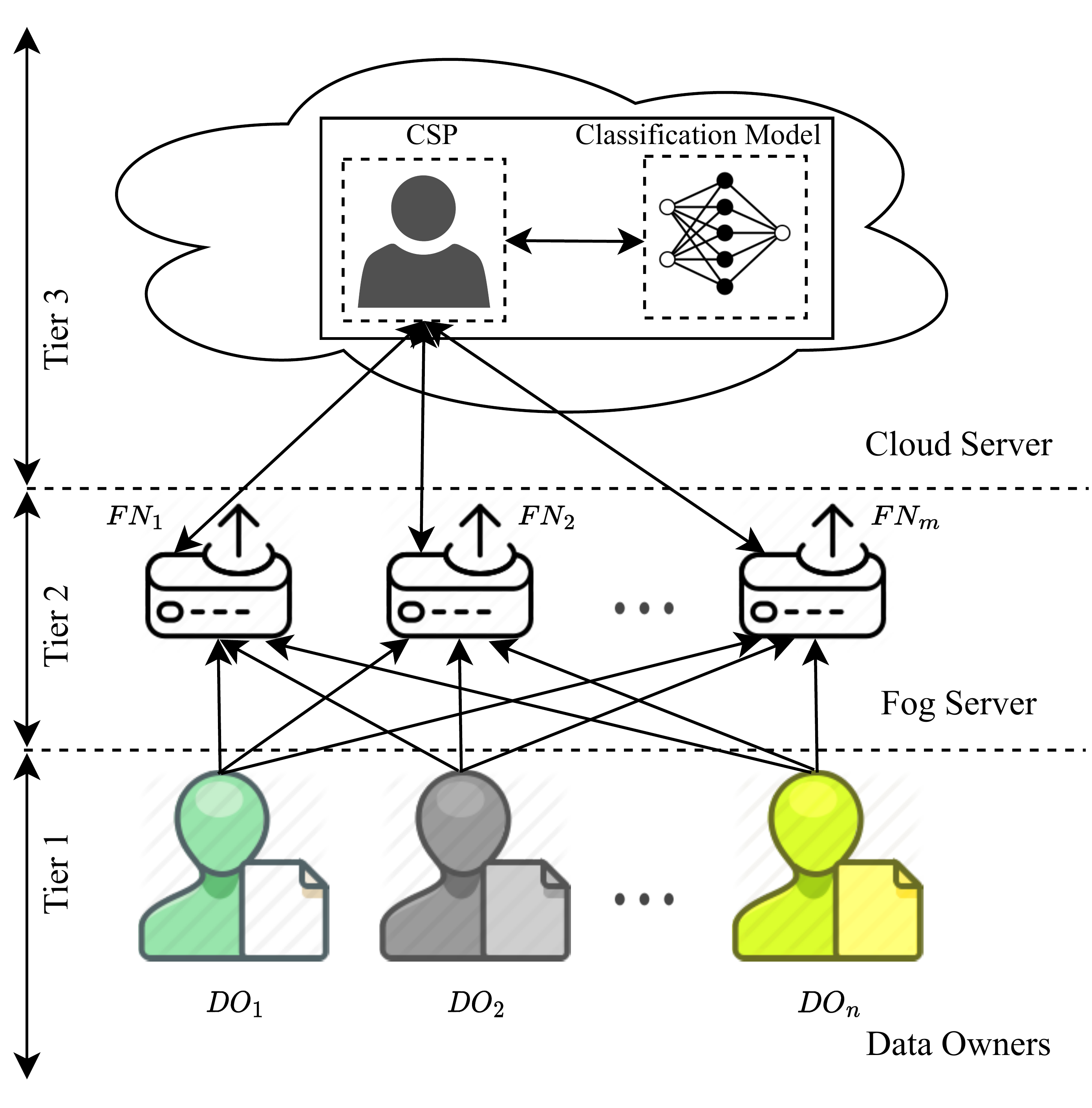}}
		\caption{System model}
		\label{fig:Systemmodel}
\end{figure*}
\begin{enumerate}
\item[a)]\textit{$Tier1$}: The bottom-most layer of the proposed architecture is tier1. At tier1, multiple data owners \{$DO_{1}$, $DO_{2}$, $\dots$, $DO_{n}$\} $\in$ $\mathbb{DO}$ exist . Each $DO_{1}$, $DO_{2}$, $\dots$, $DO_{n}$ has its data  \{$D_{1}$, $D_{2}$, $\dots$, $D_{n}$\} $\in$ $\mathbb{D}$, which is used for storage, computation, performing the classification task on the machine learning algorithm. $D_{1}$, $D_{2}$, $\dots$, $D_{n}$ includes sensitive data. For example, $DO_{1}$ has $D_{1}$ data containing \{$X_{11}$, $X_{12}$, $\dots$, $X_{1m}$\} feature instance, $DO_{2}$ has $D_{2}$ data containing the feature instance \{$X_{21}$, $X_{22}$, $\dots$, $X_{2m}$\}, $\dots$, $DO_{n}$ has $DO_{n}$ data containing \{$X_{n1}$, $X_{n2}$, $\dots$, $X_{nm}$\} feature instance. To preserve the privacy of data $\mathbb{D}$ from the adversary, noise vector  \{$N_{1}$, $N_{2}$, $\dots$, $N_{n}$\} $\in$ $\mathbb{N}$, is generated by the $DO_{1}$, $DO_{2}$, $\dots$, $DO_{n}$. This generated noise $N_{1}$, $N_{2}$, $\dots$, $N_{n}$ is injected into the data$D_{1}$, $D_{2}$, $\dots$, $D_{n}$, respectively. After adding noise, $DO_{1}$, $DO_{2}$, $\dots$, $DO_{n}$ have noised-added data $D_{i}^{N_{i}}$ =\{$D_{1}^{N_{1}}$, $D_{2}^{N_{2}}$, $\dots$, $D_{n}^{N_{n}}$\}. At tier1, this noise-added data is divided vertically, i.e.\{$X_{11}$, $X_{21}$, $\dots$, $X_{n1}$\}, \{$X_{12}$, $X_{22}$, $\dots$, $X_{n2}$\}, and \{$X_{1m}$, $X_{2m}$, $\dots$, $X_{nm}$\}. The vertically partitioned data is transferred to tier2.
\item[b)]\textit{$Tier2$}: The middle layer of the system model is Tier2. This is known as the fog computing layer (including routers, access points, gateways, and switches). The main component of this layer is multiple fog nodes \{$FN_{1}$, $FN_{2}$, $\dots$, $FN_{s}$\} $\in$ $FN_{j}$, where $j$ $\in$ [1, s]. These fog nodes are smart intermediate devices with data storage, computation, routing, and packet forwarding capabilities. Each fog nodes access and store the vertically partitioned data of the multiple data owners from tier1. For example, the instance feature \{$X_{11}$, $X_{21}$, $\dots$, $X_{n1}$\} is stored on the $FN_{1}$ node, \{$X_{12}$, $X_{22}$, $\dots$, $X_{n2}$\} is stored on $FN_{2}$ node, and \{$X_{1m}$, $X_{2m}$, $\dots$, $X_{nm}$\} is stored on the $FN_{s}$ node, respectively. The adversary cannot access the entire data when storing partitioned data on different fog nodes.
\item[c)]\textit{$Tier3$}: The cloud service provider is presented at the topmost tier in the system model. This tier contains many high-speed servers, data centers as well as the machine learning model. $CSP$ offers ample storage space, computing capabilities, and classification services to cloud users as per their request at the topmost tier. $CSP$ is a semi-honest entity. Besides, to perform the classification task, $CSP$ obtains the noisy data set from different fog nodes. The achieved results from the classification model, $CSP$ shares the outcomes to $DO_{1}$, $DO_{2}$, $\dots$, $DO_{n}$ through $FN_{1}$, $FN_{2}$, $\dots$, $FN_{s}$ . 
\end{enumerate}
\section{Conclusion}
The work focuses on preserving the privacy of outsourced data of multiple data owners. Conventional cloud computing technology provides many benefits as storage, computation, and machine learning tasks to the cloud user. But the cloud storage creates several security conflicts. Whenever data owners store their data for storage on the cloud and perform machine learning tasks, the owners then lose their rights and are unable to handle and maintain the physical storage of their data. To tackle the problem of privacy protection in cloud storage and access the classification service from cloud computing, we propose a PPOD scheme. This proposed scheme is based on the fog computing model and differential privacy. Using differential privacy, noise is injected by the data owners, and data is divided vertically. The separated data is stored on the various fog nodes. Therefore, adversaries cannot access the actual and whole data from one place. 
\bibliographystyle{unsrt}
\bibliography{references.bib} 
\end{document}